\RequirePackage[mathlines]{lineno} 
\documentclass[prd,twocolumn,showpacs,amsmath,amssymb,aps]{revtex4}
\usepackage{overpic,graphicx}
\usepackage{dcolumn}
\usepackage{bm}
\usepackage{rotating}
\usepackage{times}
\usepackage{amsfonts}
\usepackage{amsthm}
\usepackage{latexsym}
\usepackage{color}
\usepackage{url}
\usepackage{subfigure}
\usepackage{enumerate}

\begin{document}
\normalsize
\parskip=5pt plus 1pt minus 1pt
\uchyph=0
\lefthyphenmin=2
\righthyphenmin=2

\title{ \boldmath Measurements of the 
branching fractions for the semi-leptonic decays
$D^+_s\to
\phi e^{+}\nu_{e}$, $\phi \mu^{+}\nu_{\mu}$, $\eta \mu^{+}\nu_{\mu}$ and
$\eta'\mu^{+}\nu_{\mu}$
}
\author{\small{
M.~Ablikim$^{1}$, M.~N.~Achasov$^{9,e}$, S. ~Ahmed$^{14}$, M.~Albrecht$^{4}$, A.~Amoroso$^{53A,53C}$, F.~F.~An$^{1}$, Q.~An$^{50,a}$, J.~Z.~Bai$^{1}$, Y.~Bai$^{39}$, O.~Bakina$^{24}$
% ...
\\
\vspace{0.2cm}
(BESIII Collaboration)\\
\vspace{0.2cm} {\it
$^{1}$ Institute of High Energy Physics, Beijing 100049, People's Republic of China\\
% ...
$^{4}$ Bochum Ruhr-University, D-44780 Bochum, Germany\\
% ...
$^{9}$ G.I. Budker Institute of Nuclear Physics SB RAS (BINP), Novosibirsk 630090, Russia\\
% ...
$^{14}$ Helmholtz Institute Mainz, Johann-Joachim-Becher-Weg 45, D-55099 Mainz, Germany\\
% ...
$^{24}$ Joint Institute for Nuclear Research, 141980 Dubna, Moscow region, Russia\\
% ...
$^{39}$ Southeast University, Nanjing 211100, People's Republic of China\\
% ...
$^{50}$ University of Science and Technology of China, Hefei 230026, People's Republic of China\\
% ...
$^{53}$ (A)University of Turin, I-10125, Turin, Italy; (B)University of Eastern Piedmont, I-15121, Alessandria, Italy; (C)INFN, I-10125, Turin, Italy\\
% ...
$^{a}$ Also at State Key Laboratory of Particle Detection and Electronics, Beijing 100049, Hefei 230026, People's Republic of China\\
% ...
$^{e}$ Also at the Novosibirsk State University, Novosibirsk, 630090, Russia\\
% ...
}
\vspace{0.4cm}
}}

\begin{abstract}
By analyzing 482 pb$^{-1}$ of $e^+e^-$ collision data collected
at the center-of-mass energy $\sqrt s=4.009$ GeV with the BESIII detector,
we measure the 
branching fractions for the semi-leptonic decays $D_{s}^{+}\to
\phi e^{+}\nu_{e}$, $\phi \mu^{+}\nu_{\mu}$, $\eta \mu^{+}\nu_{\mu}$ and
$\eta'\mu^{+}\nu_{\mu}$
to be
${\mathcal B}(D_{s}^{+}\to\phi e^{+}\nu_{e})=(2.26\pm0.45\pm0.09)$\%,
${\mathcal B}(D_{s}^{+}\to\phi \mu^{+}\nu_{\mu})=(1.94\pm0.53\pm0.09)$\%,
${\mathcal B}(D_{s}^{+}\to\eta \mu^{+}\nu_{\mu})=(2.42\pm0.46\pm0.11)$\% and
${\mathcal B}(D_{s}^{+}\to\eta'\mu^{+} \nu_{\mu}) = (1.06\pm0.54\pm0.07)$\%,
where the first and second uncertainties are statistical and systematic, respectively.
The branching fractions for the three semi-muonic decays
$D_s^+\to\phi \mu^+\nu_\mu, \eta \mu^+\nu_\mu$ and $\eta' \mu^+\nu_\mu$
are determined for the first time and
that of $D^+_s\to \phi e^+\nu_e$
is consistent with the world
average value within uncertainties.
\end{abstract}
\maketitle

\section{\boldmath Introduction}
The semi-leptonic (SL) decays of charmed mesons
($D^{0(+)}$ and $D^+_s$) provide an ideal window to
explore heavy quark decays,
as the strong and weak effects can be well separated in theory.
The Operator Product Expansion (OPE) model predicts that
the partial widths of the inclusive SL decays of $D^{0(+)}$
and $D^+_s$ mesons should be equal,
up to 
non-factorizable components~\cite{OPE}.
However, the CLEO collaboration reported
a deviation 18\% for inclusive partial
widths between $D^{0(+)}$ and $D^+_s$ SL decays,
which is more than 3 times
of the experimental uncertainties~\cite{prd81}.
Ref.~\cite{ISGW2} argues that
this deviation may be due to that
the spectator quark masses $m_u$ and $m_s$ differ on the
scale of the daughter quark mass $m_s$ in the Cabibbo-favored SL transition.
Therefore, comprehensive or improved  measurements
of the 
branching fractions (BFs) for the exclusive SL
decays of $D^{0(+)}$ and $D^+_s$ will benefit the understanding of this difference.
Also, these measurements can serve to
verify the theoretical calculations on these decay rates.

In recent years, the $D^{0(+)}$ SL decays have been well studied
with good precision~\cite{pdg2014}.
However, the progress in the studies of
the $D^+_s$ SL decays is still relatively slow.
Up to now,
only $D^+_s$ semi-electronic decays have been
investigated by various experiments~\cite{prd_80_052007, barbar1, arx_1505_04205, BESetaev} and no
measurements of $D^+_s$ semi-muonic
decays have been reported.
We here report the first measurements of the 
BFs of the semi-muonic decays $D^+_s\to \eta\mu^+\nu_\mu$,
$\eta^\prime\mu^+\nu_\mu$ and $\phi\mu^+\nu_\mu$
as well as a measurement of the BF of the
semi-electronic decay $D^+_s\to \phi e^+\nu_e$. Charge-conjugate
decays are implied throughout this paper,
unless otherwise stated.
Among them, the studies of $D^+_s \to \eta^{(\prime)}\mu^+\nu_\mu$
may also shed light on $\eta-\eta^\prime-$glueball mixing~\cite{Glueball},
as their decay rates are expected to be sensitive to the
$\eta-\eta^\prime$ mixing angle~\cite{MixAngle}.

In this paper, all measurements are preformed by analyzing the same data set
as used in our previous measurements of
$D^+_s \to \eta^{(\prime)} e^+\nu_e$~\cite{BESetaev}.
This data set, corresponding to an integrated luminosity
of 482 pb$^{-1}$~\cite{data}, was
collected at the center-of-mass energy $\sqrt s=4.009$ GeV
with the BESIII detector.

\section{\boldmath BESIII and monte carlo}
BESIII is a cylindrical spectrometer that is
composed of a Helium-gas based main drift chamber
(MDC), a plastic scintillator time-of-flight (TOF) system,
a CsI (Tl) electromagnetic calorimeter (EMC), a
superconducting solenoid providing a 1.0 T magnetic field
and a muon counter in the iron flux return yoke of the magnet. The
momentum resolution of charged tracks in the MDC is 0.5\% at a transverse momentum of
1 GeV/$c$, and the photon energy resolution is 2.5\%(5.0\%) at
an energy of 1 GeV in the barrel (endcap) of the EMC.
More details about BESIII detector are described in Ref.~\cite{bes3}.

A GEANT4-based ~\cite{GEANT4} Monte Carlo (MC)
simulation software, which includes the geometric description of
the BESIII detector and its response, is used to determine
detection efficiencies and estimate background contributions.
The simulation is implemented by the MC event generator KKMC~\cite{KKMC}
using EvtGen~\cite{EvtGen,EvtGen1}, including the beam energy spread and the
effects of initial-state radiation (ISR)~\cite{ISR}. Final-state
radiation (FSR) of the charged tracks is simulated with
the PHOTOS package~\cite{PHOTOS}.
An inclusive MC sample corresponding to an integrated
luminosity of 11~fb$^{-1}$ is generated at $\sqrt s=4.009$ GeV,
which includes open charm production,
ISR production of low-mass vector charmonium states,
continuum light quark production, $\psi(4040)$ decays and QED events.
The open charm processes are simulated with cross sections
taken from Ref.~\cite{CrossS}.
The known decay modes of the charmonium states are produced by
EvtGen with the BFs quoted from the Particle Data Group (PDG)~\cite{pdg2014},
and the unknown decay modes are simulated by the LundCharm generator~\cite{LundCharm}.
The SL decays of interest are simulated incorporating
with the ISGW2 form-factor model~\cite{ISGW2}.

\section{\boldmath Data analysis}

In $e^{+}e^{-}$ collisions at $\sqrt s=$ 4.009 GeV,
$D^+_s$ and $D^-_s$ mesons can only be produced jointly without additional hadrons.
Thus in an event where a $D^{-}_s$ meson (called single-tag (ST)
$D^{-}_s$ meson) is fully reconstructed, the presence of a $D^{+}_s$ meson is
guaranteed. In the systems recoiling against the ST $D^-_s$ mesons, we can
select the SL decays of interest (called double-tag (DT) events).
For a specific ST mode $i$,
the observed yields of ST ($N^i_{\rm ST}$) and DT ($N^i_{\rm DT}$)
are given by
\begin{equation}
N^{i}_{\rm ST} = 2 N_{D_s^+D_s^-} {\mathcal B}^i_{\rm ST} \epsilon^i_{\rm ST}
\label{eq:1}
\end{equation}
and
\begin{equation}
N^{i}_{\rm DT} = 2 N_{D_s^+D_s^-} {\mathcal B}^i_{\rm ST} {\mathcal B}_{\rm SL}
\epsilon^i_{\rm DT},
\label{eq:2}
\end{equation}
respectively. Where $N_{D_s^+D_s^-}$ is the total number of $D_s^+D_s^-$ pairs produced in data,
${\mathcal B}^i_{\rm ST}$ and ${\mathcal B}_{\rm SL}$
are the BFs for the ST mode $i$ and the SL decay of interest,
$\epsilon^i_{\rm ST}$ is the efficiency of reconstructing the ST mode $i$ (called the ST efficiency),
and $\epsilon^i_{\rm DT}$ is the efficiency of simultaneously
finding the ST mode $i$ and the SL decay (called the DT efficiency).
The efficiency of ST and DT are determined by MC simulation.
In this analysis, the ST $D_{s}^{-}$
mesons are reconstructed in ten hadronic decay modes:
$K^+K^-\pi^{-}$,
$\phi\rho^-$,
$K^{0}_{S}K^{+}\pi^{-}\pi^{-}$, $K^{0}_{S}K^{-}\pi^{+}\pi^{-}$,
$K^{0}_{S}K^{-}$, $\pi^+\pi^-\pi^-$, $\eta\pi^{-}$,
$\eta'_{\eta\pi^+\pi^-}\pi^{-}$,
$\eta'_{\gamma\rho^0}\pi^{-}$ and $\eta\rho^-$.
Candidates of $K^0_S$, $\pi^0$, $\eta$, $\phi$, $\rho^-$,
$\eta'_{\eta\pi^+\pi^-}$ and $\eta'_{\gamma\rho^0}$ are selected using
$K^0_S\to \pi^+\pi^-$, $\pi^0\to\gamma\gamma$, $\eta\to\gamma\gamma$,
$\phi\to K^+K^-$, $\rho^-\to\pi^0\pi^-$,
$\eta'\to\pi^+\pi^-\eta$ and $\eta'\to\gamma\rho^0$ decays, respectively.
The ST modes are selected separately according to their charges.
Based on Eq.~(\ref{eq:1}) and Eq.~(\ref{eq:2}), the 
BF of the SL decay
can be determined according to
\begin{equation}
{\mathcal B}_{\rm SL} = \frac{N^{\rm tot}_{\rm DT}}{N^{\rm tot}_{\rm ST}\bar \epsilon_{\rm SL}},
\label{eq:3}
\end{equation}
by considering the multiple ST modes,
where $N_{\rm DT}^{\rm tot}$ and $N_{\rm ST}^{\rm tot}$
are the total yields of ST and DT events for multiple ST modes,
$\bar \epsilon_{\rm SL} =
\sum_i (N^i_{\rm ST} \epsilon^i_{\rm DT}/\epsilon^i_{\rm ST})/N^{\rm tot}_{\rm ST}$
is the weighted efficiency of detecting the SL decay
for the multi-ST mode
according to the yields of different ST modes.

All charged tracks are required to be within a polar-angle ($\theta$)
range of $|\cos\theta|<0.93$. The good charged tracks,
except for those from $K^0_S$ decays, are required to
originate within an interaction region defined by $V_{xy}<1.0$ cm
and $|V_{z}|<10.0$ cm, where $V_{xy}$  and $|V_{z}|$
are the distances of closest approach of the reconstructed track
to the interaction point (IP) perpendicular to
the beam direction and along the beam direction, respectively.
Particle identification (PID) is implemented with
the ionization energy loss ($dE/dx$) measured by the MDC and
the time of flight recorded by the TOF.
For each charged track,
the combined confidence levels for pion and kaon hypotheses
($CL_\pi$ and $CL_K$ ) are calculated, respectively.
A pion (kaon) is identified by requiring $CL_\pi>0$ and $CL_\pi>CL_K$
($CL_K>0$ and $CL_K>CL_\pi$).
The $K^{0}_{S}$ candidates are reconstructed with two opposite charged
tracks which satisfy $|V_{\rm z}| < 20$ cm and are assumed to be pions without PID.
A vertex constrained fit is performed to the $\pi^+\pi^-$ combinations,
and the fitted track parameters are used in the further
analysis. The distance $L$ of the secondary vertex to the
IP is also required to be
positive with respect to the $K^0_S$ flight direction.
$K^0_S$ candidates are required to have $\pi^+\pi^-$ invariant mass
within $(0.485, 0.511)$ GeV/$c^2$.
Photon candidates are chosen from isolated clusters in the EMC.
The deposited energy of a neutral cluster is required to be larger than
25 MeV in the barrel region ($|\cos\theta|<0.80$) or
50 MeV in the end-cap region ($0.86<|\cos\theta|<0.92$).
The angle between the photon candidate and the nearest charged
track should be larger than $10^\circ$.
To suppress electronic noise and energy deposits unrelated to the events,
the difference between the EMC time and the event start time
is required to be within (0, 700)~ns.
The $\pi^0$ and $\eta$ candidates are reconstructed
with $\gamma\gamma$ pair with invariant mass within
$(0.115, 0.150)$ and $(0.510, 0.570)$ GeV/$c^2$.
To improve momentum resolution,
a kinematic fit is performed to constrain the $\gamma\gamma$ invariant mass
to the nominal $\pi^0$ or $\eta$ mass,
and the fitted momenta of $\pi^0$ or $\eta$ are used in the further analysis.
To select candidates of $\phi$, $\rho^-$, $\eta'_{\pi^+\pi^-\eta}$
and $\eta'_{\gamma\rho^0}$ mesons,
the invariant masses of $K^+K^-$, $\pi^-\pi^0$,
$\pi^+\pi^-\eta$ and $\gamma\rho^0$ are required to be within
$(1.005, 1.040)$, $(0.570, 0.970)$,
$(0.943, 0.973)$ and  $(0.932, 0.980)$ GeV/$c^2$, respectively.
For $\eta'_{\gamma\rho^0}$ candidate, the $\pi^+\pi^-$ invariant mass is
additionally required to fall in $(0.570,0.970)$ GeV/$c^2$
to reduce combinatorial backgrounds.

\begin{figure*}[htbp]
\includegraphics[width=18cm]{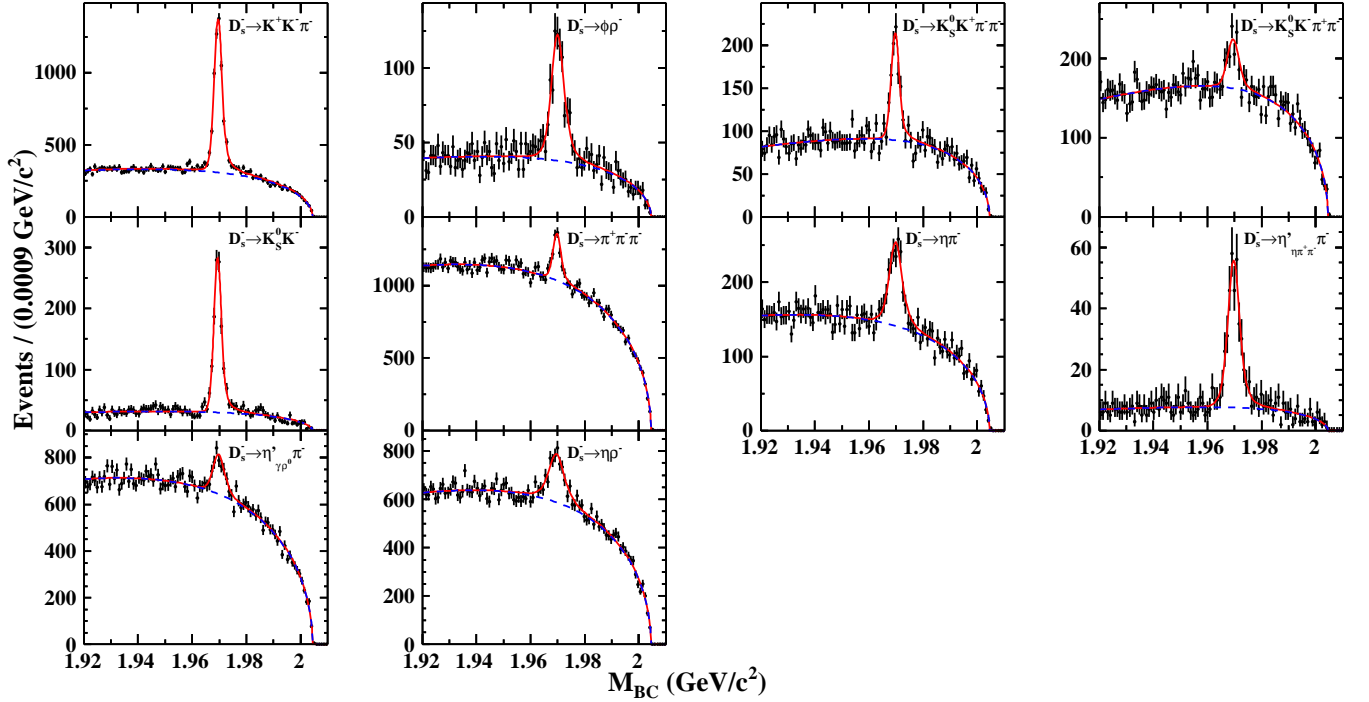}
 \caption{ \label{fig:Stag_Mbc}
  (Color online) Fits to the $M_{\rm BC}$ distributions of the
  ST $D_{s}^{-}$ decay modes.
  The dots with error bar are data,
  the red solid curves represent the total fits, and
  the blue dashed curves describe the fitted backgrounds.
  }
\end{figure*}

The ST $D_{s}^{-}$ meson is identified
using the energy difference $\Delta E \equiv E_{D^-_s}-E_{\rm beam}$ and
the beam-constrained mass $M_{\rm BC} \equiv \sqrt{E^2_{\rm beam}-|\vec{p}_{D^-_s}|^{2}}$,
where $E_{\rm beam}$ is the beam energy,
$E_{D^-_s}$ and $|\vec{p}_{D^-_s}|$
are the total energy and momentum
of the ST $D^-_s$ candidate in the $e^+e^-$ center-of-mass frame.
For each ST mode, only the one with the minimum $|\Delta E|$ is retained
if there are multiple combinations in an event.
To suppress combinatorial backgrounds,
modes dependent $\Delta E$ requirements,
which correspond to $(-3.0,+3.0)$ times of the resolution
around the fitted $\Delta E$ peak,
are imposed on the ST $D^-_s$ candidates.
Figure~\ref{fig:Stag_Mbc} shows the $M_{\rm BC}$ distributions
of $D_s^-$ candidates for individual ST mode.
To obtain the ST yield ($N^i_{\rm ST}$),
we perform a maximum likelihood fit on these $M_{\rm BC}$ distributions.
In the fits, we use the MC-simulated signal shape convoluted with
a Gaussian function to model the $D^-_s$ signals
and an ARGUS function~\cite{argus} to describe the combinatorial backgrounds.
The events with $M_{\rm BC}$ within
a mass window of $(-4.0,+5.0)$ times of the resolution
around the $D_s^-$ nominal mass~\cite{pdg2014} (called $M_{\rm BC}$ signal region)
are kept for further analysis.
For each ST mode,
the ST yield is obtained by integrating
the $D_{s}^{-}$ signal over the corresponding $M_{\rm BC}$ signal region.
The ST efficiency for the individual mode ($\epsilon^i_{\rm ST}$)
is determined by analyzing the inclusive MC sample.
Table~\ref{tab:SingleTag} summarizes
the requirements on $\Delta E$ and $M_{\rm BC}$,
the ST yields in data and the ST efficiencies.
The total ST yield ($N^{\rm tot}_{\rm ST}$) is $13092\pm247$.

\begin{table*}[hbtp]
\caption{\label{tab:SingleTag}
Summary of the requirements on $\Delta E$ and $M_{\rm BC}$,
the ST yields in data ($N_{\rm ST}$)
and the ST efficiencies ($\epsilon_{\rm ST}$), which do not include the BFs for
daughter particles $\pi^0$, $K^{0}_{S}$, $\phi$, $\eta$ and $\eta'$
for the individual ST mode.
The uncertainties are statistical only.
}
\renewcommand\arraystretch{1.1}
\begin{tabular}{lcccc} \hline\hline
ST mode & $\Delta E$ (GeV) & $M_{\rm BC}$ (GeV/$c^2$) & $N^i_{\rm ST}$ & $\epsilon^i_{\rm ST}$ (\%) \\ \hline
$D^-_s\to K^+K^-\pi^-$                      &$(-0.020,0.017)$&$(1.9635,1.9772)$&$4820\pm95 $&$39.95\pm0.09$ \\
$D^-_s\to \phi\rho^-$                       &$(-0.036,0.023)$&$(1.9603,1.9820)$&$619 \pm39 $&$10.88\pm0.07$ \\
$D^-_s\to K^{0}_{S}K^+\pi^-\pi^-$           &$(-0.018,0.014)$&$(1.9632,1.9781)$&$581 \pm40 $&$24.05\pm0.17$ \\
$D^-_s\to K^{0}_{S}K^-\pi^+\pi^-$           &$(-0.016,0.012)$&$(1.9621,1.9777)$&$400 \pm50 $&$22.51\pm0.22$ \\
$D^-_s\to K^{0}_{S}K^-$                     &$(-0.019,0.020)$&$(1.9640,1.9761)$&$1065\pm38 $&$46.89\pm0.21$ \\
$D^-_s\to \pi^+\pi^-\pi^-$                  &$(-0.026,0.022)$&$(1.9624,1.9787)$&$1500\pm125$&$54.35\pm0.19$ \\
$D^-_s\to \eta\pi^-$                        &$(-0.052,0.058)$&$(1.9599,1.9823)$&$834 \pm56 $&$48.36\pm0.27$ \\
$D^-_s\to \eta'_{\eta\pi^+\pi^-}\pi^-$       &$(-0.025,0.024)$&$(1.9602,1.9814)$&$325 \pm22 $&$23.47\pm0.22$ \\
$D^-_s\to \eta'_{\gamma\rho^0}\pi^-$         &$(-0.041,0.033)$&$(1.9611,1.9803)$&$1110\pm106$&$37.11\pm0.18$ \\
$D^-_s\to \eta\rho^-$                       &$(-0.058,0.041)$&$(1.9576,1.9844)$&$1838\pm120$&$26.11\pm0.10$ \\
\hline
Total &  &  & $13092\pm247$ &  \\
\hline\hline
\end{tabular}
\end{table*}

The SL decays $D^+_s\to
\phi e^{+}\nu_{e}$, $\phi \mu^{+}\nu_{\mu}$,
$\eta \mu^{+}\nu_{\mu}$ and $\eta' \mu^{+}\nu_{\mu}$
are selected recoiling against the ST $D_{s}^{-}$ mesons.
The charge of the electron (muon) candidate
is required to be opposite to that of the ST $D_{s}^{-}$ meson.
For electron (muon) PID, the $dE/dx$, TOF and EMC information is used
to form the combined confidence levels for electron, muon,
pion and kaon hypotheses ($CL_e$, $CL_\mu$, $CL_\pi$ and $CL_K$).
The electron candidates  should satisfy
$CL_e/(CL_e+CL_{\pi}+CL_K)>0.8$ and $CL_e>0.001$,
while the muon candidates are required
$CL_\mu>CL_{e}$, $CL_\mu > CL_K$ and $CL_\mu>0.001$.
It is required that there is no extra charged track except for
those used in the DT event selection.
For $D^+_s\to \eta^{(\prime)}\mu^+\nu_\mu$ decays,
the energy deposited in the EMC by muon is required
to be less than 300 MeV and the maximum energy
($E_{\rm extra\gamma}^{\rm max}$) of the extra photons,
which are not used in the DT event selection, is required to be less than 200 MeV.

\begin{figure}[htbp]
\includegraphics[width=8cm]{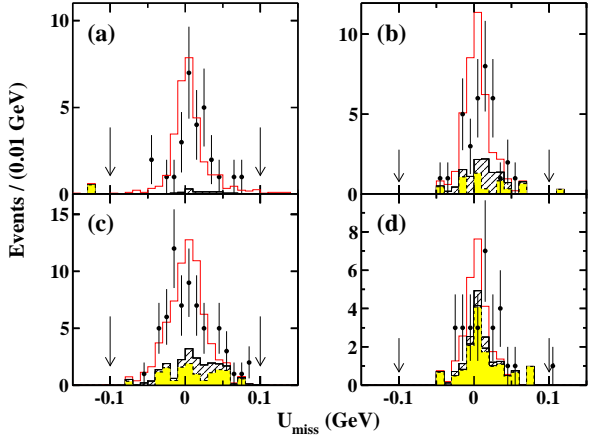}
\caption{ \label{fig:Umiss_etaev_data_Inc}
Distributions of $U_{\rm miss}$ of the candidate events for
$D_{s}^{+}\to$
(a) $\phi e^{+}\nu_{e}$,
(b) $\phi \mu^{+}\nu_{\mu}$,
(c) $\eta \mu^{+}\nu_{\mu}$ and
(d) $\eta'\mu^{+}\nu_{\mu}$
where the pair of arrows represent the signal region.
The dots with error bar are data,
the red histograms are inclusive MC,
and the yellow and oblique-line hatched histograms represent
the scaled `real-$D^-_s$' and `non-$D^-_s$' backgrounds.
}
\end{figure}

The undetected neutrino in the SL decay is inferred by a kinematic variable
$U_{\rm miss} \equiv E_{\rm miss}-|\vec p_{\rm miss}|$,
where $E_{\rm miss} \equiv \sqrt s - \sum_j E_j$ is
the missing energy
and $\vec p_{\rm miss} \equiv -\sum_j\vec p_j$ is the missing momentum.
Here, the index $j$ runs over all the particles
used in the DT event selection,
$E_j$ and $\vec p_j$
are the energy and momentum of the $j$-th particle in the $e^+e^-$ rest frame.
The $U_{\rm miss}$
distribution of
the SL decays candidates is expected to peak near zero.
To further suppress backgrounds from the hadronic decays
$D^+_s\to \phi(\eta,\eta')\pi^+$ and $\phi(\eta,\eta')\pi^+\pi^0$
for semi-muonic decays,
we define a variable
$\delta  E=E_{\rm beam}-(E_{\phi(\eta,\eta')}+E_{\mu^+~{\rm as}~\pi^+}+E_{\nu_\mu~{\rm as}~\pi^0})$,
where $E_{\phi(\eta,\eta')}$ is the energy of $\phi(\eta,\eta')$
candidate,
$E_{\mu^+~{\rm as}~\pi^+}$ is the energy of $\mu^+$
candidate by assuming it is pion, and
$E_{\nu_\mu~{\rm as}~\pi^0}$ is the energy of missing particle by assuming to be $\pi^0$
(calculated with $\vec{p}_{\rm miss}$).
The DT candidate events are required 
to have $\delta E$ within
$(-0.080,-0.010)$, $(-0.100,0)$, $(-0.070,-0.015)$ and $(-0.060,-0.015)$ GeV
for $D^+_s\to\phi\mu^+\nu_\mu$, $\eta\mu^+\nu_\mu$, $\eta'_{\eta\pi^+\pi^-}\mu^+\nu_\mu$
and  $\eta'_{\gamma\rho^0}\mu^+\nu_\mu$, respectively.
Figure~\ref{fig:Umiss_etaev_data_Inc} shows the $U_{\rm miss}$
distributions of the accepted candidate events for the SL decays in data.
The $U_{\rm miss}$ signal region is defined as $(-0.10, 0.10)$~GeV,
in which we observe
$28.0\pm5.3$, $34.0\pm5.8$, $64.0\pm8.0$ and $28.0\pm5.3$
candidate events for
$D_{s}^{+}\to\phi e^{+}\nu_{e}$,
$\phi\mu^{+}\nu_{\mu}$,
$\eta\mu^{+}\nu_{\mu}$,
and $\eta'_{\eta\pi^+\pi^- \rm and \gamma\rho^0}\mu^{+}\nu_{\mu}$,
respectively.

Some background events may also survive the selection
criteria of the SL decays of interest.
The backgrounds can be classed into two categories.
Those background events, in which the ST $D^-_s$ meson is reconstructed correctly but
the SL decay is mis-identified,
are defined as `real-$D^-_s$' background.
The other background events, in which the ST $D^-_s$ meson is reconstructed incorrectly,
are called as `non-$D^-_s$' background.
The number of `real-$D^-_s$' background events is estimated by analyzing
the inclusive MC sample. While the `non-$D^-_s$' background yield is evaluated
by using the events of data within the $M_{\rm BC}$ sideband region, which
is defined
to be $(1.920, 1.950)$ and
$(1.990, 2.000)$~GeV/c$^2$ on the $M_{\rm BC}$ distribution. The background yield
in the $M_{\rm BC}$ sideband region is then scaled
by the ratio of the background integral areas
between the $M_{\rm BC}$ signal and sideband regions.

\begin{table*}[hbtp]
\begin{center}
\caption{\label{tab:Results_Data}
The numbers used to extract the BFs of SL decay as well as the resultant BFs.
The uncertainties are statistical only.
}
\renewcommand\arraystretch{1.2}
\begin{tabular}{lccccc} \hline\hline
Decay mode   & $N^{\rm obs}_{\rm DT}$ & $N^{\rm bkg}_{{\rm real}-D^-_s}$ & $N^{\rm bkg}_{{\rm non}-D^-_s}$ &
$\bar \epsilon_{\rm SL}$ (\%)& $\mathcal B_{\rm SL}$ (\%)\\\hline
$D_s^+\to\phi e^+\nu_e$&$28.0\pm5.3$ &$1.6\pm0.2$ & $0.0^{+0.1}_{-0.0}$& $18.2\pm0.1$&$2.26\pm0.45$\\
$D_s^+\to\phi \mu^+\nu_\mu$&$34.0\pm5.8$ &$6.8\pm0.5$ &$5.1\pm1.6$ & $17.8\pm0.1$&$1.94\pm0.53$\\
$D_s^+\to\eta \mu^+\nu_\mu$&$64.0\pm8.0$&$7.0\pm0.5$&$12.6 \pm2.6$ & $35.6\pm0.2$&$2.42\pm0.46$\\
$D_s^+\to\eta'\mu^+\nu_\mu$&$28.0\pm5.3$ & $3.7\pm0.4$& $14.0\pm2.6$& $16.2\pm0.1$& $1.06\pm0.54$\\ \hline\hline
\end{tabular}
\end{center}
\end{table*}

The DT yields observed in data ($N^{\rm obs}_{\rm DT}$),
the expected number of `real-$D^-_s$' and `non-$D^-_s$' background
($N^{\rm bkg}_{{\rm real}-D^-_s}$ and
$N^{\rm bkg}_{{\rm non}-D^-_s}$) as well as
the weighted efficiencies of detecting the SL decays according to the ST yields of data
($\bar \epsilon_{\rm SL}$) are summarized in Table~\ref{tab:Results_Data},
where the efficiencies $\bar \epsilon_{\rm SL}$
do not include the BFs
of $\phi$, $\eta$ and $\eta'$ in the SL decays.
So, the 
BFs for the SL decays are determined by
\begin{equation}
{\mathcal B}_{\rm SL} = \frac{N^{\rm obs}_{\rm DT}-N^{\rm bkg}_{{\rm real}-D^-_s}-N^{\rm bkg}_{{\rm non}-D^-_s}}{N^{\rm tot}_{\rm ST}\bar \epsilon_{\rm SL} {\mathcal B}_{\rm sub}},
\label{eq:4}
\end{equation}
where ${\mathcal B}_{\rm sub}$ denotes the BFs for the
daughter particles $\phi$, $\eta$ and $\eta'$ quoted from PDG~\cite{pdg2014}.
Inserting the numbers of $N^{\rm obs}_{\rm DT}$,
$N^{\rm bkg}_{{\rm real}-D^-_s}$,
$N^{\rm bkg}_{{\rm non}-D^-_s}$,
$N_{\rm ST}^{\rm tot}$,
$\bar \epsilon_{\rm SL}$ and
${\mathcal B}_{\rm sub}$ in Eq.~(\ref{eq:4}),
we obtain the 
BFs for $D_{s}^{+}\to\phi e^{+}\nu_{e}$,
$\phi \mu^{+}\nu_{\mu}$,
$\eta \mu^{+}\nu_{\mu}$
and $\eta' \mu^{+}\nu_{\mu}$,
respectively.
These results are summarized in Table \ref{tab:Results_Data}.

\section{\boldmath Systematic uncertainties}
In the 
BF measurements using DT method,
the systematic uncertainties arising from the ST selection are almost canceled.
Main systematic uncertainties in the measurements for BFs of SL decays
are discussed below.
\begin{table*}[hbtp]
\begin{center}
\caption{\label{tab:sys_tot}
Systematic uncertainties (in \%) in the BF measurements.
The sources tagged with `$^c$'
are regarded as common systematic uncertainties between
the two $\eta^\prime$ decay modes.}
\renewcommand\arraystretch{1.2}
\begin{tabular}{lccccc} \hline\hline
Source &$D_s^+\to \phi e^{+}\nu_{e}$ &
$D_s^+\to \phi \mu^{+}\nu_{\mu}$ &
$D_s^+\to \eta \mu^{+}\nu_{\mu}$ &
$D_s^+\to \eta'_{\eta\pi^+\pi^-}\mu^{+}\nu_{\mu}$ &
$D_s^+\to \eta'_{\gamma\rho^0}\mu^{+}\nu_{\mu}$ \\
\hline
ST yield                                      & 1.8    & 1.8    & 1.8   & 1.8$^{c}$    & 1.8$^{c}$ \\
Tracking for $K^+$ ($\pi^+$)                  & 2.0    & 2.0    &---    & 2.0$^{c}$    & 2.0$^{c}$   \\
PID for $K^+$ ($\pi^+$)                       & 2.0    & 2.0    & ---   & 2.0$^{c}$    & 2.0$^{c}$   \\
Tracking for $e^+~(\mu^+)$                    & 1.0    & 1.0    & 1.0   & 1.0$^{c}$    & 1.0$^{c}$ \\
PID for $e^+~(\mu^+)$                         & 0.9    & 2.4    & 1.5   & 1.9$^{c}$    & 1.9$^{c}$ \\
$E_{\rm extra \gamma}^{\rm max}$ requirement  & ---    & ---    & 2.5   & 2.5$^{c}$    & 2.5$^{c}$   \\
$\phi (\eta,\eta')$ reconstruction            & 0.4    & 0.4    & 2.3   & 2.5    & 2.8   \\
$\delta E$ requirement                        & ---    & 0.7    & 1.2   & 1.7    & 1.8 \\
Background subtraction                        & 0.2    & 1.5    & 1.2   & 3.1    & 3.0 \\
MC statistics                                 & 0.5    & 0.6    & 0.4   & 0.6    & 0.6   \\
MC model                                      & 1.4    & 1.1    & 0.7   & 2.5    & 2.2   \\
BFs of $\phi$ and $\eta(')$                            & 1.0    & 1.0    & 0.5   & 1.6    & 1.7 \\
\hline
Total                                         & 4.0    & 4.8    & 4.7   & 7.0    & 7.1 \\
\hline \hline
\end{tabular}
\end{center}
\end{table*}

\begin{enumerate}[a.]
\item{\it ST yield.}
The uncertainty of the total ST yield is estimated to be 1.8\%
by comparing the integrated and counted ST yields
(calculated by subtracting the background yields from total events
without performing a fit)
in the $M_{\rm BC}$ signal region.
\item{\it Tracking and PID.}
The uncertainties in the tracking and PID for charged kaon and
pion are investigated
with the control sample of DT hadronic $D\bar D$ events and are
assigned to be 1.0\% and 1.0\% per track individually.
The efficiencies of the tracking and PID for electron and muon are
studied by varying with the polar angle $\cos\theta$ and momentum
with the control samples $e^+e^-\to\gamma e^+e^-$ and
$e^+e^-\to\gamma \mu^+\mu^-$ events, respectively.
These efficiencies are weighted
according to $\cos\theta$ and momentum distributions
of the electron and muon in the SL decays. The resultant differences of the
two-dimensional weighted tracking and PID efficiencies for electron and muon
between data and MC simulation are regarded as the relevant uncertainties.
\item
{\it $E_{\rm extra\gamma}^{\rm max}$ requirement.}
The efficiency of $E_{\rm extra\gamma}^{\rm max}$ requirement
is investigated with fully reconstructed DT hadronic
decays $\psi(4040)\to D^*\bar{D}+c.c.$.
The difference of the efficiencies with the requirement of
$E_{\rm extra \gamma}^{\rm max}< 200$~MeV
between data and MC simulation is found to be $(1.9\pm0.6)$\%.
To be conservative, we assign 2.5\% to be the associated systematic uncertainty.
\item
{\it $\phi$ ($\eta$, $\eta'$) reconstruction.}
The reconstruction efficiencies
for the $\phi$, $\eta$ and $\eta'$ candidates,
which include the mass window requirement and photon selection,
are estimated with the control samples of
$D^+\to\phi\pi^+$, $D^0\to K_S^0\eta$,
$D^0\to K_S^0\eta'_{\pi^+\pi^-\eta}$ and
$K_S^0\eta'_{\gamma\rho^0}$, respectively.
The differences of
efficiencies between data and MC simulation are estimated to be 0.4\%, 2.3\%, 2.5\% and 2.8\%
for $\phi$, $\eta$, $\eta'_{\pi^+\pi^-\eta}$ and $\eta'_{\gamma\rho^0}$, respectively,
which are assigned as the associated uncertainties.
\item
{\it $\delta E$ requirement.}
The uncertainties from $\delta E$ requirements are estimated
by varying the $\delta E$ requirements
by $\pm10\%$. The changes of the BFs, which are
0.7\%, 1.2\%, 1.7\% and 1.8\% for
$D_{s}^{+}\to\phi\mu^{+}\nu_{\mu}$, $\eta \mu^{+}\nu_{\mu}$,
$\eta'_{\eta\pi^+\pi^-} \mu^{+}\nu_{\mu}$
and $\eta'_{\gamma\rho^0} \mu^{+}\nu_{\mu}$, respectively,
are taken as the corresponding uncertainties.
\item
{\it Background subtraction.}
Two aspects uncertainties associated
with background subtraction are considered separately.
The `${\rm real}-D^-_s$' background is estimated with the inclusive MC samples, thus,
we vary the quoted BFs of the main background sources
$D_{s}^{+}\to\phi\mu^{+}\nu_{\mu}$, $\phi \rho^+$, $\eta \rho^+$,
$\eta'_{\eta\pi^+\pi^-}\rho^+$ and $\eta'_{\gamma\rho^0} \rho^+$
by $1\sigma$ quoted in PDG~\cite{pdg2014}.
The `${\rm non}-D^-_s$' background is estimated with the candidate events
in the $M_{\rm BC}$ sideband region. We then
shift the $M_{\rm BC}$
sideband region by $\pm5$ MeV/$c^2$.
The quadratic sum of these two effects on the measured BFs,
which are 0.2\%, 1.5\%, 1.2\%, 3.1\% and 3.0\% for
$D^+_s\to \phi e^+\nu_e$, $\phi \mu^+\nu_\mu$, $\eta \mu^+\nu_\mu$,
$\eta'_{\eta\pi^+\pi^-}\mu^+\nu_\mu$ and $\eta'_{\gamma\rho^0} \mu^+\nu_\mu$, respectively,
are treated as the systematic uncertainties.
\item
{\it MC statistics.}
The uncertainties in the weighted efficiencies
are mainly due to limited MC statistics, which
are 0.5\%, 0.6\%, 0.4\%, 0.6\% and 0.6\%
for $D_{s}^{+}\to\phi e^+ \nu_e$,
$\phi\mu^{+}\nu_{\mu}$,
$\eta \mu^{+}\nu_{\mu}$,
$\eta'_{\eta\pi^+\pi^-} \mu^{+}\nu_{\mu}$
and $\eta'_{\gamma\rho^0} \mu^{+}\nu_{\mu}$, respectively.
The effects of the statistical uncertainty of ST yields
of data is negligible for the weighting efficiencies.
\item
{\it MC model.}
The uncertainty associated with MC model is studied with an
alternative SL form-factor model,
$i.e.$,
the simple pole model~\cite{POLE}.
The resultant differences on DT efficiencies with respect to the nominal values,
which are 1.4\%, 1.1\%, 0.7\%, 2.5\% and 2.2\%
for $D_{s}^{+}\to\phi e^+ \nu_e$,
$\phi\mu^{+}\nu_{\mu}$,
$\eta\mu^{+}\nu_{\mu}$,
$\eta'_{\eta\pi^+\pi^-} \mu^{+}\nu_{\mu}$
and $\eta'_{\gamma\rho^0} \mu^{+}\nu_{\mu}$, respectively,
are considered as the associated systematic uncertainties.
\item
{\it BFs of $\phi$ and $\eta(')$.}
The BFs for $\phi\to K^+K^-$, $\eta\to\gamma\gamma$, $\eta'\to\eta\pi^+\pi^-$
and $\eta'\to\gamma\rho^0$ are quoted from the PDG~\cite{pdg2014}.
Their uncertainties are 1.0\%, 0.5\%, 1.6\% and 1.7\%, respectively.
\end{enumerate}

The individual systematic uncertainties discussed above are summarized in
Table \ref{tab:sys_tot} and the total systematic uncertainties are the quadratic sum
of the individual ones.
The sources tagged with `$^c$' are common systematic uncertainties between
the two $\eta^\prime$ decay modes and the other sources are independent.
Finally, we assign 7.1\% as the total systematic uncertainty for
$D^+_s\to \eta' \mu^+ \nu_\mu$.

\section{\boldmath Summary}

In summary, by analyzing the 482 pb$^{-1}$ data collected at
$\sqrt s=4.009$ GeV with the BESIII detector,
we determine the 
BFs for the SL decays
$D_{s}^{+}\to\phi e^{+}\nu_{e}$, $\phi \mu^{+}\nu_{\mu}$,
$\eta \mu^{+}\nu_{\mu}$ and $\eta'\mu^{+} \nu_{\mu}$.
Table \ref{tab:summary_BFs} presents the comparisons of the measured
BFs with the world average values.
The BFs of the semi-muonic decays $D_s^+\to \phi\mu^+\nu_\mu$,
$\eta\mu^+\nu_\mu$ and $\eta'\mu^+\nu_\mu$
are determined for the first time
and are compatible with those of
the corresponding semi-electronic decays~\cite{pdg2014}.
The BF of $D^+_s\to \phi e^+\nu_e$ agrees with
the world average value~\cite{pdg2014} within uncertainties.
And, the results are consistent with previous experimental measurements
and support that the ratio of $D^+_s$ and $D^{0(+)}$ differs from unity,
there is an indication of difference
between $D^{0(+)}$ and $D^+_s$ meson SL
decay widths~\cite{prd81}.
Combining the previous BESIII measurements for semi-electronic decays~\cite{BESetaev}
and this work, we calculate the ratios between the semi-electronic and semi-muonic decays,
to be ${\mathcal B}(D^+_s\to \phi\mu^+\nu_\mu)/{\mathcal B}(D^+_s\to \phi e^+\nu_e)= 0.86\pm0.29$,
${\mathcal B}(D^+_s\to \eta\mu^+\nu_\mu)/{\mathcal B}(D^+_s\to \eta e^+\nu_e)= 1.05\pm0.24$ and
${\mathcal B}(D^+_s\to \eta\mu^+\nu_\mu)/{\mathcal B}(D^+_s\to \eta e^+\nu_e)= 1.14\pm0.68$
individually, where most of systematic uncertainties are canceled.
The ratios are consistent with 1 within the uncertainties,
and no obvious lepton universality violation is observed.
Moreover, the ratio of
${\mathcal B}(D^+_s\to \eta \mu^+\nu_\mu)$ over ${\mathcal B}(D^+_s\to \eta' \mu^+\nu_\mu)$
is calculated to be $0.44\pm0.23$, which is in agreement with
those of previous measurements~\cite{prd_80_052007,arx_1505_04205,BESetaev,CLEORatio}
within uncertainties and
provides complementary data to probe
the $\eta-\eta^\prime-$glueball mixing.

\begin{table*}[hbtp]
\begin{center}
\caption{\label{tab:summary_BFs}
Summary of the 
BFs and comparing with the world average values~\cite{pdg2014}.
}
\renewcommand\arraystretch{1.2}
\begin{tabular}{cccccc} \hline\hline
$\mu^+$ mode                 &$\mathcal B_{\rm BESIII}$ (\%)&$\mathcal B_{\rm PDG}$ (\%)   &$e^+$ mode&$\mathcal B_{\rm BESIII}$ (\%)&$\mathcal B_{\rm PDG}$ (\%)\\\hline
$D_s^+\to\phi \mu^+\nu_\mu$&$1.94\pm0.53\pm0.09$&--&$D_s^+\to\phi e^+\nu_e$&$2.26\pm0.45\pm0.09$&$2.39\pm0.23$\\
$D_s^+\to\eta \mu^+\nu_\mu$&$2.42\pm0.46\pm0.11$&--&$D_s^+\to\eta e^+\nu_e$&$2.30\pm0.31\pm0.08$~\cite{BESetaev}&$2.28\pm0.24$\\
$D_s^+\to\eta'\mu^+\nu_\mu$&$1.06\pm0.54\pm0.07$&--&$D_s^+\to\eta'e^+\nu_e$&$0.93\pm0.30\pm0.05$~\cite{BESetaev}&$0.68\pm0.16$\\ \hline\hline
\end{tabular}
\end{center}
\end{table*}

\section{\boldmath Acknowledgments}
The BESIII collaboration thanks the staff of BEPCII and the IHEP computing center for their strong support. This work is supported in part by National Key Basic Research Program of China under Contract No. 2015CB856700; National Natural Science Foundation of China (NSFC) under Contracts Nos. 11235011, 11335008, 11425524, 11625523, 11635010, 11675200; the Chinese Academy of Sciences (CAS) Large-Scale Scientific Facility Program; the CAS Center for Excellence in Particle Physics (CCEPP); Joint Large-Scale Scientific Facility Funds of the NSFC and CAS under Contracts Nos. U1332201, U1532257, U1532258; CAS under Contracts Nos. KJCX2-YW-N29, KJCX2-YW-N45, QYZDJ-SSW-SLH003; 100 Talents Program of CAS; National 1000 Talents Program of China; INPAC and Shanghai Key Laboratory for Particle Physics and Cosmology; German Research Foundation DFG under Contracts Nos. Collaborative Research Center CRC 1044, FOR 2359; Istituto Nazionale di Fisica Nucleare, Italy; Joint Large-Scale Scientific Facility Funds of the NSFC and CAS; Koninklijke Nederlandse Akademie van Wetenschappen (KNAW) under Contract No. 530-4CDP03; Ministry of Development of Turkey under Contract No. DPT2006K-120470; National Natural Science Foundation of China (NSFC) under Contract No. 11505010; National Science and Technology fund; The Swedish Resarch Council; U. S. Department of Energy under Contracts Nos. DE-FG02-05ER41374, DE-SC-0010118, DE-SC-0010504, DE-SC-0012069; University of Groningen (RuG) and the Helmholtzzentrum fuer Schwerionenforschung GmbH (GSI), Darmstadt; WCU Program of National Research Foundation of Korea under Contract No. R32-2008-000-10155-0

\end{document}